\documentclass[12pt]{iopart}
\usepackage{graphicx}% Include figure files
\usepackage{dcolumn}% Align table columns on decimal point
\usepackage{bm}% bold math

\begin{document}

\title{Reevaluation of the $^{14}$O($\alpha$,p)$^{17}$F Resonant Reaction Rate}

\author{J.J. He$^1$, H.W. Wang$^2$, J. Hu$^{1}$, S.W. Xu$^1$}

\address{
$^1$ Institute of Modern Physics, Chinese Academy of Sciences (CAS), Lanzhou 730000, China \\
$^2$ Shanghai Institute of Applied Physics, Chinese Academy of Sciences (CAS), Shanghai 201800, China \\
}

\ead{jianjunhe@impcas.ac.cn}
\begin{abstract}
The reaction rates of the key stellar reaction of $^{14}$O($\alpha$,p)$^{17}$F have been reevaluated. It is thought that the previous 1$^-$ assignment for the 6.15-MeV
state is incorrect by a careful reanalysis of the previous experimental data [J. G\'{o}mez del Campo {\it et al.}, Phys. Rev. Lett. {\bf 86}, 43 (2001)].
Most probably, the 6.286-MeV state is the key 1$^-$ state and the 6.15-MeV state is a 2$^-$ one, and hence the resonance at $E_x$=6.286 MeV ($J^{\pi}$=1$^-$) actually
dominates the reaction rates in the temperature region of astrophysical interests. The newly calculated reaction rates for the $^{14}$O($\alpha$,p)$^{17}$F reaction
are quite different from the previous ones, for instance, it's only about 1/6 of the previous value around 0.4 GK, while it's about 2.4 times larger than the previous
value around 2 GK. The astrophysical implications have been briefly discussed based on the present conclusions.
\end{abstract}

%Uncomment for PACS numbers title message
%\pacs{00.00, 20.00, 42.10}
% Keywords required only for MST, PB, PMB, PM, JOA, JOB?
%\vspace{2pc}
%\noindent{\it Keywords}: Article preparation, IOP journals
% Uncomment for Submitted to journal title message
%\submitto{\JPA}
% Comment out if separate title page not required
\maketitle

\section{Introduction}
Explosive hydrogen and helium burning are thought to be the main source of energy generation and a source for the nucleosynthesis of heavier elements in cataclysmic
binary systems, for example, x-ray bursters, {\em etc.}~\cite{bib:woo76,bib:cha92,bib:wie98}. During an x-ray burst (a high temperature and high density astrophysical
site), Hydrogen and Helium rich material from a companion star form an accretion disk around the surface of a neutron star where the H and He transferred from the disk
begin to pile up. The $\alpha$$p$ chain is initiated through the reaction sequence
$^{14}$O($\alpha$,$p$)$^{17}$F($p$,$\gamma$)$^{18}$Ne($\alpha$,$p$)$^{21}$Na~\cite{bib:bar00}, and increase the rate of energy generation by 2 orders of
magnitude~\cite{bib:wie98}. In x-ray burster scenarios, the nucleus $^{14}$O($t_{1/2}$=71 s) forms an important waiting point, and the ignition of the
$^{14}$O($\alpha$,$p$)$^{17}$F reaction at temperatures $\sim$0.4 GK produces a rapid increase in power and can lead to breakout from the hot CNO cycles into the
$rp$-process with the production of medium mass proton-rich nuclei~\cite{bib:sch98,bib:sch01,bib:bre09}. Excepting the $^{15}$O($\alpha$,$\gamma$)$^{19}$Ne reaction,
this reaction is arguably the most important reaction to be determined for x-ray burster scenarios.

Wiescher {\em et al.}~\cite{bib:wie87} calculated the reaction rates of the $^{14}$O($\alpha$,p)$^{17}$F reaction, and shown that the resonant reaction rates dominated
the total rates above temperature ~0.4 GK. However, Funck {\em et al.}~\cite{bib:fun88,bib:fun89} found that direct-reaction contributions to the $\ell$=1 partial wave
are comparable to or even greater than the resonant contributions at certain temperatures. Because the resonant reaction rates of $^{14}$O($\alpha$,p)$^{17}$F depend
sensitively on the excitation energies, spins, and partial and total widths of the relevant resonances in $^{18}$Ne, Hahn {\em et al.}~\cite{bib:hah96} extensively
studied the levels in the compound system $^{18}$Ne~\cite{bib:hah96} by several reactions, such as, $^{16}$O($^3$He,$n$)$^{18}$Ne,$^{12}$C($^{12}$C,$^6$He)$^{18}$Ne
as well as $^{20}$Ne($p$,$t$)$^{18}$Ne reactions. Based on the firmer experimental results, they concluded that this reaction rate, in the important temperature regime
$\sim$0.5-1 GK, was dominated by reactions on a single 1$^-$ resonance at an excitation energy of 6.150 MeV lying 1.036 MeV above the $^{14}$O+$\alpha$ threshold
($Q_{\alpha}$=5.114 MeV). Harss {\em et al.}~\cite{bib:har99} studied the time reverse reaction $^{17}$F($p$,$\alpha$)$^{14}$O in inverse kinematics with $^{17}$F beam
at Argonne, and identified three levels at 7.16, 7.37, 7.60 MeV and determined their resonance strengths as well. Later, G\'{o}mez del Campo {\em et al.}~\cite{bib:gom01}
used the $p$($^{17}$F,$p$) resonant elastic scattering on a thick CH$_2$ target to look for resonances of astrophysical interest in $^{18}$Ne at ORNL. In the region
investigated, they located four resonances at excitation energies of 4.52, 5.10, 6.15, and 6.35 MeV in $^{18}$Ne, and J$^{\pi}$=1$^-$, 2$^-$ were respectively assigned
to the last two states based on their $R$-matrix analysis. Subsequently, Harss {\em et al.}~\cite{bib:har02} extracted the resonance strength and the width
$\Gamma_\alpha$ for the 6.15-MeV state based on this 1$^-$ assignment together with the excitation function obtained from their previous work~\cite{bib:har99}.
Recently, the inelastic component of this key 1$^-$ resonance in the $^{14}$O($\alpha$,p)$^{17}$F reaction has been studied by a new highly sensitive technique at
ISOLDE/CERN~\cite{bib:hjj09}, and found that this inelastic component will enhance the reaction rate, contributing approximating equally to the ground-state component
of the reaction rate, however not to the relative degree suggested in Ref.~\cite{bib:bla03}.

As a summary, all the previous discussions and calculations~\cite{bib:hah96,bib:har02,bib:bla03,bib:hjj09} related to the reaction rates of $^{14}$O($\alpha$,p)$^{17}$F
are based on the 1$^-$ assignment for the 6.15-MeV state. In this work, the spin-parities for three relevant states (at 6.15, 6.286, and 6.345 MeV) will be reassigned
by a carefully reanalyzing the experimental data measured at ORNL~\cite{bib:gom01}, and it is found that most probably the 6.286-MeV state is the key 1$^-$ state and
the 6.15-MeV state is a 2$^-$ one. Therefore the resonance at $E_x$=6.286 MeV ($J^{\pi}$=1$^-$) actually dominates the reaction rates in the temperature region of
astrophysical interests rather than the 6.15-MeV state does. The reaction rates have been recalculated and the astrophysical consequences have been briefly discussed
based on the present new assignments.

\section{Data Reanalysis}
The previous ORNL experimental data\cite{bib:gom01}, {\em i.e.}, the integrated cross sections of the $^{17}$F+$p$ elastic scattering in the angular range of
$\theta_\mathrm{CM}$=162$^\circ$$\sim$178$^\circ$, have been reanalyzed by a $R$-matrix method~\cite{bib:lan58,bib:des03,bib:bru02,bib:ang_alex}
(see example~\cite{bib:alex}) carefully.
A channel radius of $r_0$=1.25 fm ($R$=$r_0$$\times$(1+17$^{1/3}$)) appropriate for the $^{17}$F+$p$ system\cite{bib:gom01,bib:wie87,bib:hah96} has been utilized in
the present $R$-matrix calculation, where the fitting results are insensitive to the choice of radius, {\em e.g.}, the resultant $\chi^2/N$ value is
changed by less than 15\% in the region of $r_0$=1.2$\sim$1.3~fm.
The ground state spin-parity configurations of $^{17}$F and the proton are 5/2$^+$ and 1/2$^+$, respectively. Thus, there are two channel spins in the elastic
channel, {\em i.e.} $s$ = 2, 3. But for a 1$^-$ state, only one unique channel spin, $s$=2, is possible for an $\ell$=1 transfer.

\begin{figure}
\begin{center}
\includegraphics[scale=0.4]{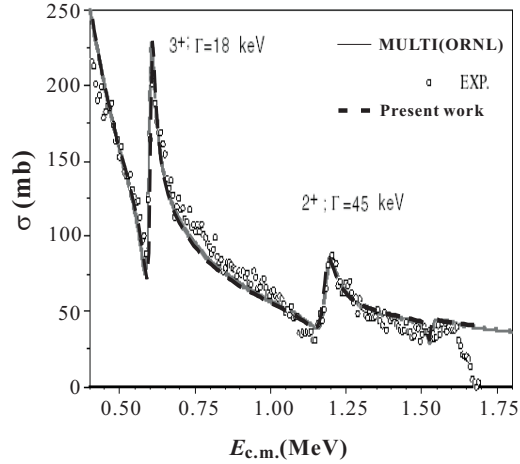}
\caption{\label{fig1} The upper panel of Fig. 2 as shown in Ref.\cite{bib:gom01}. The present $R$-matrix fit with the same resonant parameters is shown for comparison.}
\end{center}
\end{figure}

\begin{figure}
\begin{center}
\includegraphics[width=8cm]{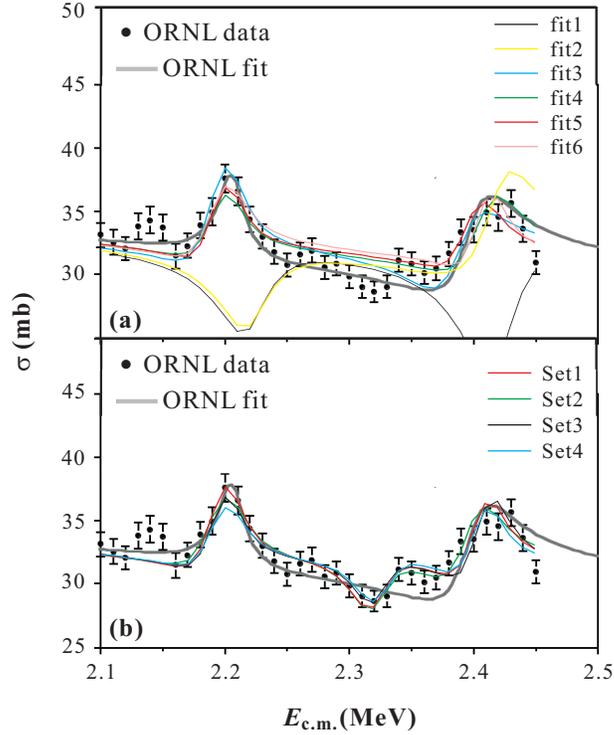}
\caption{\label{fig2} $R$-matrix fits for the experimental data measured at ORNL~\cite{bib:gom01} for the elastic scattering of $^{17}$F+$p$. The vertical scale
corresponds to the angle integrated cross sections in the range $\theta_\mathrm{CM}$=162$^\circ$$\sim$178$^\circ$. (a) fitting for two resonances, (b) fitting for
three resonances. All curves are convoluted by an assumed 10-keV energy resolution except those two labeled by `fit1' and `fit2'. For comparison, the ORNL fit is
shown as well. See text for details.}
\end{center}
\end{figure}

The fitting curve shown in the upper panel of Fig. 2 in~\cite{bib:gom01} has been exactly reproduced by utilizing their resonant parameters as shown in Fig.~\ref{fig1},
{\em i.e.}, $E_{cm}$=0.6 MeV, 3$^+$, $\Gamma$=18 keV; $E_{cm}$=1.18 MeV, 2$^+$, $\Gamma$=45 keV (1.18 MeV was mistyped by 1.118 MeV in~\cite{bib:gom01}); and the
last `groove' structure can be reproduced by the following parameters: $E_{cm}$=1.53 MeV, 2$^-$, $\Gamma$=5 keV, which is consistent with the results of
Ref.~\cite{bib:hah96}($\Gamma$$\leq$20 keV). In Fig.~\ref{fig1}, no energy resolution is convoluted in the fitting curve, and it implies that this effect was not
considered (or neglected) in the previous analysis~\cite{bib:gom01}.
The data shown in the lower panel of Fig. 2 in~\cite{bib:gom01} have been carefully analyzed, and it's found that the 1$^-$ assignment for the 6.15-MeV state is very
unlikely as shown in Fig. 2(a) (see curves labeled by `fit1' and `fit2') with G\'{o}mez del Campo {\it et al.}'s parameters listed in Table 1. Here no energy resolution
is convoluted in the curves labeled by `fit1' and `fit2'. It shows that the shape of the 1$^-$ resonance ($\ell$=1) is of a `groove' structure instead of a `bump' one.
In addition, all possible combinations of
different spin-parity assignments for these two states have been attempted and the most probable fitting curves are shown in Fig. 2(a) with parameters listed in
Table 1. In order to achieve a better fit, all curves are convoluted by an assumed 10-keV energy resolution, of course
this will not affect the spin-parity assignments for the resonances. Furthermore, we have tried to fit the data with three resonances, and the most probable fitting
curves are shown in Fig. 2(b) with parameters listed in Table 1. It's very obvious this kind
of three-resonance fits reproduce the experimental data better than those two-resonance ones, especially the `groove' structure at $E_{cm}$=2.32 MeV can be well
fitted (see $\chi^2/N$ values in Table 1).
The present $R$-matrix analysis shows that two states at $E_{cm}$=2.20($E_x$=6.12), 2.40($E_x$=6.32) both possibly have $J^{\pi}$=2$^-$ or 3$^-$, while the state at
$E_{cm}$=2.32($E_x$=6.24) most probably has $J^{\pi}$=1$^-$. These three states should correspond to the $E_x$=6.150, 6.345, and 6.286 MeV states observed
before~\cite{bib:hah96} within a $\sim$30 keV uncertainty.

\Table{\label{table1} Resonant parameters used in Fig.\ref{fig2}. Here, resonance energies ($E_r$) are in units of MeV, and proton partial widths ($\Gamma_p$) in keV.
The parameters in `fit1' and `fit2' were used in the previous work~\cite{bib:gom01}. See text for details.}
\br
 & & \centre{3}{Resonance 1} &  \centre{3}{Resonance 2} & \centre{3}{Resonance 3}\\
Sets & $\chi^2/N$ & \crule{3} & \crule{3} & \crule{3}\\
\ns
 &  & $E_{r1}$ & $J^{\pi}$[$\ell,s$] & $\Gamma_p$ & $E_{r2}$ & $J^{\pi}$[$\ell,s$] & $\Gamma_p$ & $E_{r3}$ & $J^{\pi}$[$\ell,s$] & $\Gamma_p$\\
\mr
fit1$^{\rm a}$ & 31.5 & 2.22 & 1$^-$[1, 2] & 50 & 2.42 & 2$^-$[1, 3] & 50 &      &             &     \\
fit2$^{\rm a}$ & 15.1 & 2.22 & 1$^-$[1, 2] & 50 & 2.42 & 2$^-$[1, 2] & 50 &      &             &     \\
fit3$^{\rm b}$ & 2.0  & 2.20 & 3$^-$[1, 3] & 15 & 2.39 & 3$^-$[1, 2] & 30 &      &             &     \\
fit4$^{\rm b}$ & 1.8  & 2.20 & 2$^-$[1, 2] & 15 & 2.41 & 2$^-$[1, 2] & 30 &      &             &     \\
fit5$^{\rm b}$ & 3.1  & 2.20 & 3$^-$[1, 3] & 10 & 2.40 & 2$^-$[1, 2] & 20 &      &             &     \\
fit6$^{\rm b}$ & 2.4  & 2.20 & 2$^-$[1, 2] & 20 & 2.41 & 3$^-$[1, 3] & 10 &      &             &     \\
Set1$^{\rm b}$ & 1.2  & 2.20 & 3$^-$[1, 3] & 12 & 2.40 & 2$^-$[1, 3] & 12 & 2.32 & 1$^-$[1, 2] & 15  \\
Set2$^{\rm b}$ & 1.1  & 2.20 & 2$^-$[1, 2] & 20 & 2.40 & 2$^-$[1, 2] & 20 & 2.32 & 1$^-$[1, 2] & 15  \\
Set3$^{\rm b}$ & 1.2  & 2.20 & 3$^-$[1, 3] & 10 & 2.41 & 3$^-$[1, 3] & 12 & 2.32 & 1$^-$[1, 2] & 10  \\
Set4$^{\rm b}$ & 1.1  & 2.20 & 2$^-$[1, 3] & 15 & 2.41 & 3$^-$[1, 2] & 10 & 2.33 & 1$^-$[1, 2] & 12  \\
\br
\end{tabular}
\item[] $^{\rm a}$ No energy-resolution convoluted in the fit curves of Fig.\ref{fig2}.
\item[] $^{\rm b}$ A 10-keV energy-resolution convoluted in the fit curves of Fig.\ref{fig2}.
\end{indented}
\end{table}

\section{Discussion}
In order to constrain the spin-parity assignments for these states, let's examine the well-known mirror
nucleus $^{18}$O, which has only three known levels $J^{\pi}$=1$^-$,(2$^-$), and 3$^-$ around this energy region~\cite{bib:hah96}.
According to the previous results~\cite{bib:hjj09,bib:bla03}, the 6.15 MeV state can decay to the first excited state (1/2$^+$) in $^{17}$F with an appreciable width.
In this case, two channel spins are $s$=0, 1, respectively. Such a decay is possible only for $\ell$=3 but, the barrier penetrability is extremely too small to produce
such a large decay width. Thus, this 6.15 MeV state most probably is a 2$^-$ state since the 1$^-$ assignment has been excluded in the present $R$-matrix analysis.
The results from the
$^{12}$C($^{12}$C,$^6$He)$^{18}$Ne and $^{20}$Ne($p$,$t$)$^{18}$Ne reactions suggest that 6.286-MeV state is of natural parity and 6.345-MeV state of unnatural parity.
Therefore, we propose that these two states most probably have 1$^-$ and 2$^-$, respectively. Actually Funck {\em et al.}~\cite{bib:fun88} predicted a 1$^-$ state at
6.294 MeV. The ORNL experimental data can be reproduced very well with these assignments (see fitting curve labeled by `Set2' in Fig. 2(b)).

Therefore only one natural-parity state, {\em i.e.}, $E_x$=6.286 ($J^{\pi}$=1$^-$, $\ell_\alpha$=1), is needed to calculate the reaction rates of
$^{14}$O($\alpha$,p)$^{17}$F in the temperature region interesting for x-ray burster scenarios~\cite{bib:hah96,bib:har02,bib:hjj09},
Its $\Gamma_\alpha$ partial width is about 13.4 eV according to the relationship of $\Gamma_\alpha \propto C^2S_\alpha \times P_\ell$($E_r$)~\cite{bib:hjj091}, while
it was 0.34 eV in the previous work~\cite{bib:hah96}; its resonant strength $\omega\gamma$$_{(\alpha,p)}$ is about 40 eV rather than the previous 2.4 eV.

\begin{figure}
\begin{center}
\includegraphics[width=8cm]{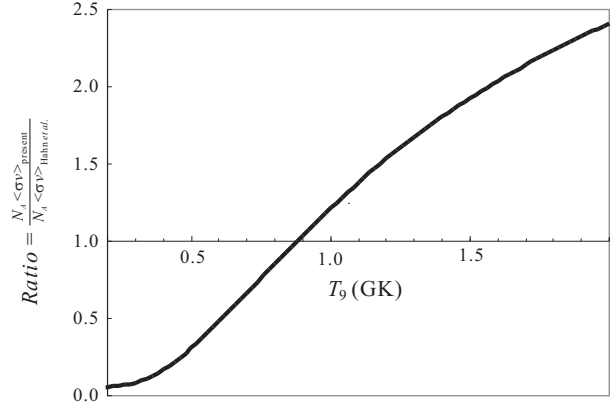}
\caption{\label{fig3} The ratios between the present resonant reaction rates and those previous ones at certain temperature range. See text for details.}
\end{center}
\end{figure}

The resonant reaction-rate ratios between the present results and the previous ones~\cite{bib:hah96} are plotted in Fig.~\ref{fig3}. Only one resonance
($E_x$=6.286 MeV) is included in the present work while two resonances ($E_x$=6.150, 6.286 MeV) involving in the previous work.
It can be seen that the present reaction rate is quite different
from the previous one, for instance, it's only about 1/6 of the previous value around 0.4 GK but, it's about 2.4 times larger than the previous value around 2 GK.
According to the present analysis, we think the 1$^-$ assignment for the 6.286-MeV state is reasonable, and hence the spin-parity assignments for the 6.150,
6.345 MeV states are rather unimportant in calculating the reaction rates ({\em i.e.}, whether they are $J^{\pi}$=2$^-$,3$^-$, or vice versa).
The present rates confirms that the $^{14}$O($\alpha$,p)$^{17}$F reaction is rather unlikely to be dominant component in the hot CNO cycles in novae environments
(instead $^{15}$O($\alpha$,$\gamma$)$^{19}$Ne reaction is, see discussions in~\cite{bib:hah96}). Due to the present rate enhancements above 0.9 GK, this reaction can,
however, contribute strongly to the breakout from the hot CNO cycle under the more extreme conditions in x-ray bursters. The present conclusion could probably affect
the onset temperature where the $\alpha$-capture dominates $\beta$-decay and a breakout from the hot CNO cycle via $^{14}$O($\alpha$,p)$^{17}$F reaction begins to take
place~\cite{bib:wie87}. However, it should be noted that the statistics of ORNL data is not very good, and hence it prevents us from putting the present
assignments on a very firm ground. Therefore more precise experiment, {{\em e.g.}}, ($p$,$t$) transfer reaction, is strongly suggested to confirm these results in the
future.

\ack
This work is financially supported by the the ``100 Persons Project"(BR091104) and the ``Project of Knowledge Innovation Program"  of Chinese Academy of Sciences
(KJCX2-YW-N32), the National Natural Science Foundation of China (10975163,10505026), and the Major State Basic Research Development Program of China (2007CB815000).

\end{document}